\documentclass[aps,preprint,onecolumn,eqsecnum,nofootinbib]{revtex4}

\usepackage{epsfig}
\usepackage{graphicx}
\usepackage{dcolumn}
\usepackage{amsmath}
\usepackage{enumerate}
\usepackage{color}

\DeclareMathOperator*{\mineta}{\min_{\eta}}

\newcommand{\capdef}{}
\newcommand{\mycaption}[2][\capdef]{\renewcommand{\capdef}{#2}%
       \caption[#1]{{\footnotesize #2}}}

\begin{document}

\vskip 1pt \hfill {IPPP/09/27}
\vskip 1pt \hfill {DCPT/09/54}

\vskip 40pt 

\title{
Measuring neutrino mass with radioactive ions in a storage ring
\vspace*{1cm}}

\author{\bf Mats Lindroos$^1$, Bob McElrath$^{2}$, Christopher
  Orme$^{1,3}$ and Thomas Schwetz$^{4}$\vspace*{1cm}} 

\affiliation{$^1$ AB division, CERN, Geneva, Switzerland} 

\affiliation{$^2$ Theory division, CERN, Geneva, Switzerland}

\affiliation{$^3$ IPPP, Department of Physics, Durham University,
  Durham DH1 3LE, United Kingdom}

\affiliation{$^{4}$ Max-Planck-Institute for Nuclear Physics, PO Box 103980, 
  69029 Heidelberg, Germany\vspace*{0.5cm}}

\begin{abstract}
  \vspace*{0.3cm} We propose a method to measure the neutrino mass
  kinematically using beams of ions which undergo beta decay. The idea
  is to tune the ion beam momentum so that in most decays, the electron
  is forward moving with respect to the beam, and only in decays near
  the endpoint is the electron moving backwards. Then, by counting the
  backward moving electrons one can observe the effect of neutrino mass
  on the beta spectrum close to the endpoint. In order to reach
  sensitivities for $m_\nu < 0.2$ eV, it is necessary to control the ion
  momentum with a precision better than $\delta p/p < 10^{-5}$, identify
  suitable nuclei with low $Q$-values (in the few to ten keV range),
  and one must be able to observe at least $\mathcal{O}(10^{18})$
  decays.
\end{abstract}

\maketitle


\section{Introduction}

There is now convincing evidence that neutrinos have mass owing to the
confirmation of neutrino oscillations. Data from
atmospheric~\cite{atm,SKatm}, solar~\cite{sol,SKsolar,SNO},
reactor~\cite{CHOOZ,PaloVerde,KamLAND} and long-baseline
experiments~\cite{K2K,MINOS} are well described by three-flavor
neutrino oscillations, with the following values for the oscillation
parameters~\cite{current}:
\begin{alignat}{1}
\vert \Delta m_{31}^{2}\vert &=2.40^{+0.12}_{-0.11}\times 10^{-3}\:\: \rm{eV}^{2} 
    \qquad \quad \sin^{2}\theta_{23}=0.50^{+0.07}_{-0.06} \nonumber\\
\Delta m_{21}^{2}&=7.65^{+0.23}_{-0.20}\times 10^{-5}\:\: \rm{eV}^{2} 
    \qquad \quad \sin^{2}\theta_{12}=0.30^{+0.01}_{-0.02} \label{E:current}
\end{alignat}
where $\Delta m_{ji}^{2}=m_{j}^{2}-m_{i}^{2}$ are neutrino
mass-squared differences, and $\theta_{ij}$ are the lepton mixing
angles. The third mixing angle is only bounded from above with
current data:
\begin{equation}
\sin^{2}\theta_{13}< 0.040 \qquad (2\sigma) \,.
\end{equation}
Neutrino oscillation experiments have yet to determine the sign of
$\Delta m_{31}^{2}$ and so we are left with two possible neutrino mass
orderings. The case $\Delta m_{31}^{2}>0$ is referred to as
\emph{`normal mass ordering'} while the case $\Delta m_{31}^{2}<0$ is 
labelled \emph{`inverted mass ordering'}. In addition, the absolute mass
scale is presently unknown. If the absolute scale is small there can be
a pronounced hierarchy between the masses dependent on the ordering. In
particular we can have
\begin{enumerate}
\item \emph{Normal mass hierarchy (NH)}: \qquad $m_{1} \ll m_{2}\ll m_{3}$
\item \emph{Inverted mass hierarchy (IH)}: \qquad $m_{3}\ll m_{1} \lesssim m_{2}$
\end{enumerate}
If the absolute scale is large compared to the mass squared
splittings, the spectrum is referred to as \emph{`quasi-degenerate'
(QD)}, where $m_{1}\simeq m_{2}\simeq m_{3}=m_{0}$ with
$m_{0}^{2}\gg\vert \Delta m_{31}^{2}\vert$. If the lightest neutrino
mass is not negligible and of the order $\sqrt{\vert\Delta
m_{31}^{2}\vert}$, then there exists a partial hierarchy.

At present there are three approaches to determining the neutrino mass
scale: through the measurement of the electron energies near the
endpoint of a beta-decay spectrum, measurement of the half-life in a
neutrino-less double beta-decay (if neutrinos are Majorana particles),
and via study of cosmological data; see Ref.~\cite{Bilenky:2002aw} for
a review. An incomplete list of phenomenological studies of
observables sensitive to absolute neutrino masses can be found in
Refs.~\cite{Feruglio:2002af, Fogli:2004as, Pascoli:2005zb,
  deGouvea:2005hj, Hannestad:2007tu, Fogli:2008ig, Maneschg:2008sf}.
Both the observation of neutrino-less double beta decay and cosmology
are model dependent approaches. In particular, it is not known whether
neutrino-less double beta decay exists, and if were observed, there
could be other mechanisms such as a $SU(2)_W$ triplet
Higgs~\cite{Mohapatra:1981pm}, Leptoquarks~\cite{Hirsch:1996ye}, and
R-parity violation in Supersymmetry~\cite{Hirsch:1995ek}.

The present bound on the neutrino mass from kinematic studies of beta decay 
endpoints was obtained from the Mainz and Troitsk Tritium beta decay experiments
\cite{Kraus:2004zw,Lobashev:1999tp,Lobashev:2001uu}:
\begin{equation}
    m^{\rm eff}_\nu < 2.3 \:\:\rm{eV} \,,
\end{equation}  
where $m^{\rm eff}_\nu$ is the effective neutrino mass in beta decay
(see below).  The next iteration of the Tritium beta decay technologies
is the KATRIN experiment, which should be able to place a limit
$m^{\rm eff}_\nu < 0.2$~eV~\cite{Osipowicz:2001sq}. However it is
unlikely that these approaches will scale to lower neutrino masses.
They can only make a positive identification of the neutrino mass if
the hierarchy is quasi-degenerate. Rhenium
calorimeters, such as proposed by the MARE
collaboration~\cite{Monfardini:2005dk}, are also
expected to achieve a limit $m^{\rm eff}_\nu < 0.2$~eV.

Thus, new experimental technologies are required to reach the
level of hierarchical masses. Furthermore, it will be important to
verify the mass-mixing hypothesis for neutrino flavor conversion, as
there are non-mass proposals which can cause neutrino 
mixing~\cite{Kostelecky:2003cr,Kostelecky:2003xn}.
This means measuring the kinematic effect of not
only the absolute mass but the mass differences as well.  This
requires precision at the $\sqrt{\Delta m_{31}^2} \simeq 0.05$~eV, or
even at $\sqrt{\Delta m_{21}^2} \simeq 0.009$~eV level.  One new
approach towards the neutrino mass, using ultra-cold atoms has been
proposed recently in~\cite{Jerkins:2009tc}.

Here we present an idea to search for
beta-decay spectrum endpoint distortions using radioactive ion
beams.  By tuning the boost factor of the ions,
only electrons very close to the endpoint of the
beta spectrum move in the direction opposite to the beam
direction in the laboratory frame. In principle, this allows one to 
search for kinematic effects
of a non-zero neutrino mass by counting the electrons with backward 
trajectories. We
explore this possibility by performing preliminary sensitivity
estimates. We specify the most important requirements on the setup in
order to achieve sensitivities below 0.2 eV.

Radioactive ion beams are currently being considered as a possible
source of neutrinos for a future long baseline neutrino oscillation
experiment~\cite{Zucchelli:2002sa}. These ``beta beams'' are subject
of intense phenomenological and R\&D studies.  Beta beams with small
ion boost factors, $\gamma \sim 10$, have been discussed in the
neutrino literature for cross section and nuclear physics measurements
at low energies, whilst boosts for long baseline experiments typically
range $\gamma = 80 - 650$.  See~\cite{Volpe:2006in, LM} for reviews
and references. The setup proposed here has many aspects in common
with such beta beams, though the boost factors needed in our case are
very close to one: $\gamma \approx 1 + Q/m_e$ or $v/c \sim
\sqrt{2Q/m_e}$, where the $Q$-value of the decay is assumed to be
small compared to the electron mass: $Q \ll m_e$. In this work we do
not propose a specific experimental scheme for the a measurement of
neutrino mass. Whether such an experiment can be integrated in a
``high-$\gamma$'' beta beam facility is an interesting question to be
addressed in future studies.

This proposal has significant challenges that must be considered.  These
include the identification of ions with low $Q$-values in order to
maximize the effect of the neutrino mass ($Q$ in the range of few to
ten~keV), as well as the required number of useful decays of order
$\gtrsim 10^{18}$. Furthermore, the momentum spread of the ions in the
beam has to be less than $10^{-5}$. This can be achieved either with
classical cooling techniques or by exploring the use of
``crystallized beam'' technology~\cite{crystallized, schramm04}.

This paper is structured as follows. In section~\ref{S:pheno} we
briefly review the phenomenology of the measurement of the neutrino
mass using endpoint studies.  In section~\ref{S:concept} we outline
our experimental proposal, while in section~\ref{S:analysis} we
present simulations of the precision this setup can achieve, and the
requirements on various ingredients to reach $m^{\rm eff}_\nu < 0.2$ eV. In
section~\ref{S:technology} we comment on challenges of the proposed
measurement. A summary follows in section~\ref{S:summary}. In
Appendix~\ref{S:R} we provide supplementary information on our
measurement strategy.


\section{Beta decay endpoint phenomenology}
\label{S:pheno}

The standard approach for a direct mass measurement is the
analysis of the endpoint region of a beta
decay~\cite{Kraus:2004zw, Lobashev:1999tp, Lobashev:2001uu}, see
Ref.~\cite{Otten:2008zz} for a review. A measurement is made by
reconstructing of the electron spectrum of the decays of ion
$I$,
\begin{equation}
I \longrightarrow I' + e^{-}+\bar{\nu}_{e} \,.
\end{equation}
It should be noted that such a measurement is sensitive to the mass scale of the
anti-neutrino which is assumed to be identical to its neutrino
counterpart owing to CPT invariance, but it need not 
be~\cite{Kostelecky:2003cr}.
Nuclei which decay through positron emission also have competing
electron capture decay channels. Electron capture dominates for
proton-rich nuclei with low $Q$-values.  Positron decay is kinematically
forbidden for $Q < 2m_{e}$ effectively prohibiting the measurement of the
neutrino mass by this method. 

The three mixing angles and CP-phase parameterize a mixing matrix, with
elements $U_{\alpha i}$, which relate the neutrino mass eigenstates with the
eigenstates that participate in the weak interaction. This mixing of the
neutrino mass eigenstates means that the spectrum of the electrons from
the decay should be considered as the incoherent sum of the spectra
associated with each neutrino mass eigenstate, 
\begin{equation}
\frac{d\Gamma}{dE_{\beta}}=\sum_{i}\vert U_{ei}\vert^{2}
\frac{d\Gamma_{i}}{dE_{\beta}},
\label{E:elecspec1}
\end{equation}
where
\begin{equation}
\frac{d\Gamma_{i}}{dE_{\beta}}=p_{\beta}E_{\beta}(E_{\rm max}-E_{\beta})
\sqrt{(E_{\rm max}-E_{\beta})^{2}-m_{i}^{2}\:\:}
\:F(Z,E_{\beta})\:S(E_{\beta})\:[1+\delta_{R}(Z,E_{R})]
\label{E:elecspec}
\end{equation}
are the individual electron spectra and $m_{i}$ is the mass for 
eigenstate $i$. A recent evaluation of the
spectrum can be found in~\cite{Gardner:2004ib, Masood:2007rc}. Here
$E_\beta$ is the kinetic energy of the electron and $E_{\rm max}$ is the
maximum electron kinetic energy for zero neutrino mass $m_i = 0$
(``endpoint energy''). $E_{\rm max}$ is given by
\begin{equation}
    \label{E:Emax}
E_{\rm max} = \frac{M^2 - {M'}^2 + m_e^2}{2M} - m_e \approx Q \,,
\end{equation}
where $M$ and $M'$ are the masses of the mother and daughter ions $I$
and $I'$, respectively, and we define the $Q$-value to be $Q \equiv M -
M' - m_e$.\footnote{Strictly speaking the $Q$-value is defined as the
mass difference of neutral atoms. Here we denote with $M$ and $M'$ the
actual masses of the ion before and after the beta decay, and
therefore we explicitly include the electron mass in the expression
for $Q$.  This holds up to corrections of order of the binding energy
of the electron.} The approximate
relation $E_{\rm max} \approx Q$ holds under the assumptions $(M - M')/M \ll 1$ and
$m_e \ll M$. Further, $S(E_{\beta})$ is a form factor that contains the
nuclear matrix element and constants
\begin{equation}
S(E_{\beta})=G_{F}^{2}\left(\frac{m_{e}^{5}c^{4}}{2\pi^{3}\hbar^{7}}\right)\: 
\cos\theta_{c}\:|\mathcal{M}_{\beta}(E_\beta)|^{2}.
\end{equation}
The Fermi function $F(Z,E_{\beta})$ describes the Coulomb
interactions of the ejected particle. For nuclear radius $R=1.2A^{1/3}$
and the definitions $\eta=\alpha Z E_{\beta}/p_{\beta}$ and $\gamma =
(1-(\alpha Z)^{2})^{1/2}$, it takes the form
\begin{equation}
F(Z,E_{\beta})=4\left(\frac{2p_{e}R}{\hbar}\right)^{2\gamma-2}\exp(\pi\eta)
\frac{\left|\Gamma(\gamma+i\eta)\right|^{2}}
{\left|\Gamma(2\gamma+1)\right|^{2}} \approx 
\frac{2\pi\eta}{1-\exp(-2\pi\eta)} \,.
\end{equation}
The Coulomb correction increases the decay rate for $\beta^{-}$ decays
and decreases it for $\beta^{+}$ decays because the attraction to the
nucleus enhances the decays with low momentum.  $\delta_{R}$ is the
contribution from electromagnetic radiative corrections
\cite{Gardner:2004ib} which are usually negligible.

The electron spectrum in Eq.~\ref{E:elecspec1} is parameterized by
the 3 neutrino masses and the mixing angles $\theta_{12},\theta_{13}$.
However, for experiments with resolution worse than $\sqrt{|\Delta
  m^2_{31}|}$, we may parameterize the spectrum with a single
effective mass $m_\nu^{\rm eff}$, see
e.g.,~\cite{Vissani:2000ci,Farzan:2001cj,Farzan:2002zq}:
\begin{equation}\label{E:effective}
    m_\nu^{\rm eff}=\sqrt{\sum_{i}\vert U_{ei}\vert^{2}\:m_{i}^{2}} \simeq
\left\{ \begin{array}{l@{\qquad}l}
    m_{\rm min}
    & + \mathcal{O}(\sqrt{\Delta m^2_{21}}) \quad
\mbox{(normal ordering)} \\[2mm]
\sqrt{m_{\rm min}^2 + |\Delta m^2_{31}|} & 
+ \mathcal{O}(\sqrt{\Delta m^2_{21}})
\quad \mbox{(inverted ordering)}
\end{array}\right.
\end{equation}
where $m_{\rm min}$ is the mass of the lightest neutrino, and the
approximate expressions hold up to terms of order $\sqrt{\Delta
m^2_{21}} \sim \sin\theta_{13}\sqrt{|\Delta m^2_{31}|} \sim
0.01$~eV. Then the spectrum becomes
\begin{equation}\label{E:eff_spect}
\frac{d\Gamma}{dE_{\beta}}\propto p_{\beta}E_{\beta}(E_{\rm max}-E_{\beta})
\sqrt{(E_{\rm max}-E_{\beta})^{2}-(m^{\rm eff}_\nu)^{2}\:\:}
\:F(Z,E_{\beta}) \,.
\end{equation}

\begin{figure}
\begin{center}
\includegraphics[width=0.7\textwidth]{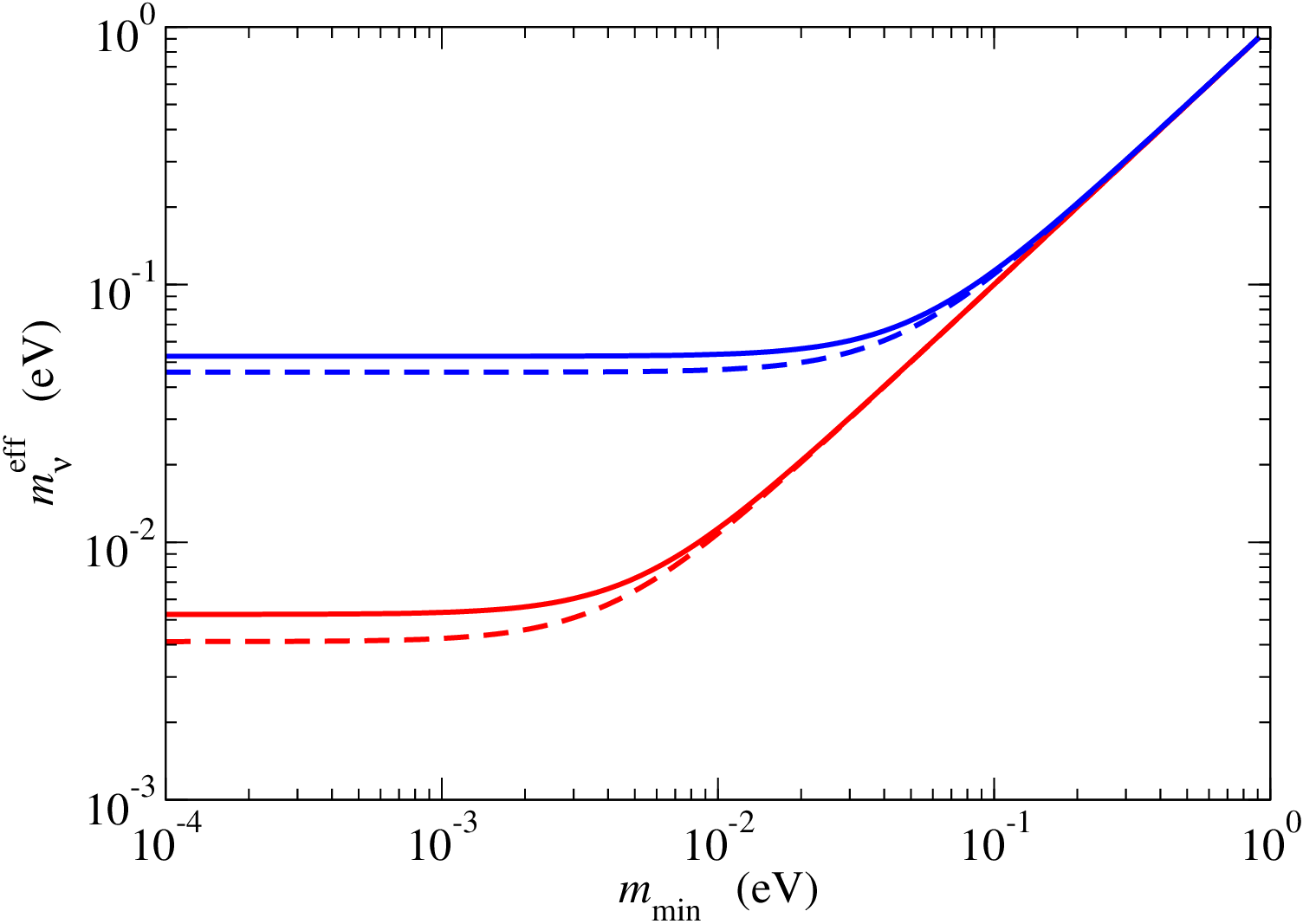}
\end{center}
\mycaption{The effective mass as a function of lightest neutrino mass
  $m_{\rm{min}}$ for normal mass ordering (red) and inverted mass
  ordering (blue). In all cases, $\theta_{13}=0$. The solid lines have
  been simulated for the $3\sigma$ upper bound for the solar and
  atmospheric parameters as defined in~\cite{current}, whilst the
  dashed lines use the $3\sigma$ lower bounds.
  }\label{Fi:effmass}
\end{figure}

In Fig.~\ref{Fi:effmass}, we show the effective mass $m_\nu^{\rm eff}$ as a
function of the lightest neutrino mass $m_{\rm min}$. The behavior
for normal mass ordering is shown in red whilst the blue lines are for
inverted mass ordering. For a small minimum neutrino mass, the
eigenstates separated by the solar splitting determine the size of the
effective mass. For normal ordering, this pair is positioned at 
$m_{\rm min}$ resulting in a small effective mass. For inverted mass
ordering, however, the solar pair is separated from $m_{\rm min}$ by
the atmospheric mass splitting. The effective mass is a factor 5
larger as a consequence. For $m_{\rm min}\gtrsim \sqrt{\Delta m_{\rm
atm}^{2}}=0.049$ eV, the atmospheric splitting is not dominant and the
effective mass does not discriminate between the two orderings.

The goal of future absolute neutrino mass experiments is to push the
sensitivity below 0.04 eV.  Failure to measure the neutrino mass above
this level identifies the mass hierarchy to be normal, on the
assumption of neutrino mixing~\cite{Bilenky:2006zd}. Failure to
measure neutrino mass above $m^{\rm eff}_\nu \sim 0.006$~eV would be
inconsistent with our current understanding of neutrino mass and
mixing given in Eq.~\ref{E:current}.
While reaching such sensitivities would be the ultimate goal of
neutrino mass measurements, in this work we are slightly less ambitious.
We have in mind experiments with sensitivities in the region $0.04
\,{\rm eV} \lesssim m^{\rm eff}_\nu \lesssim 0.2 \,\rm eV$, and hence we can
describe the spectrum by Eq.~\ref{E:eff_spect} using the
effective neutrino mass $m^{\rm eff}_\nu$. 

From the expression for the spectrum in Eq.~\ref{E:eff_spect} it
becomes clear that the effect of the neutrino mass is larger for
decays with a small $Q$-value. Most previous and present experiments
are using Tritium, with a relatively low endpoint energy of 18.6~keV
and a half life of 12.3~y. Alternatively, the MARE project uses
$^{187}$Re, with an even smaller endpoint of 2.47~keV, at the price of
a much longer half life of 43.2~Gy. In this work we do not choose a
specific ion. Instead we perform the analysis as a function of the
$Q$-value. The identification of a suitable ion is central if the
approach proposed in this paper is to be realized.


\section{The Concept}
\label{S:concept}

\begin{figure}
\begin{center}
\includegraphics[width=0.7\textwidth]{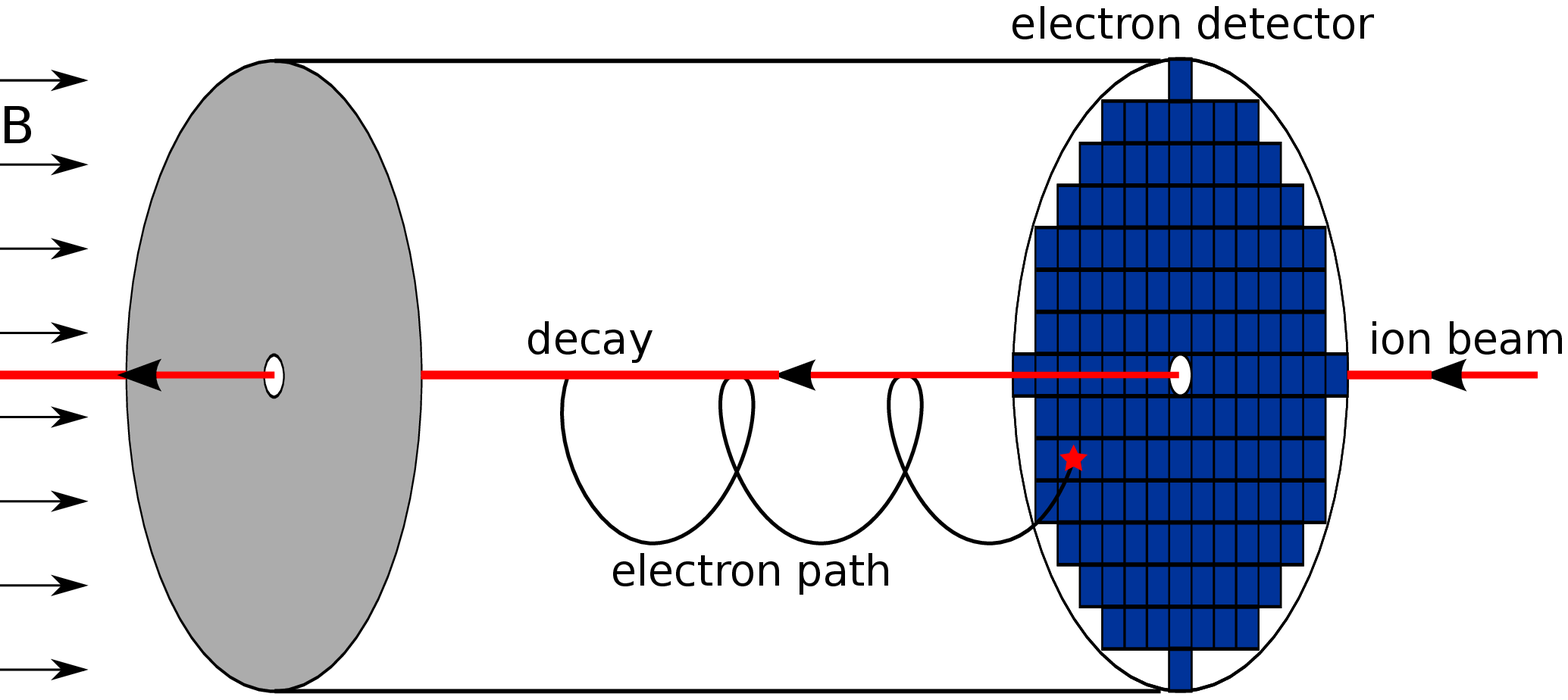}
\end{center}
\mycaption{Diagram of the proposed experiment.  The ion beam enters
  an evacuated cavity whose back wall holds an electron detector. Each
  ejected electron follows a helical trajectory. Electrons moving in
  the backward direction in the laboratory frame are counted by the
  detector. See also Fig.~\ref{Fi:R}.  }
\label{Fi:diagram}
\end{figure}

In this article we consider the possibility of using a very low boost
ion beam as a tool to measure the neutrino mass scale, by observing
backward moving electrons in the laboratory frame. A low boost 
($v/c \sim \sqrt{2Q/m_e}$) radioactive ion beam is sent through an
evacuated chamber 
with a weak magnetic field parallel to the beam line. A detector is set up on
the back wall of the chamber to record the number of electrons still traveling
backwards after the boost, see Fig.~\ref{Fi:diagram}. Therefore, the purpose 
of the boost is to perform a cut on the electron momenta, only selecting 
electrons very close to the spectrum endpoint. 

Let a boosted ion have a velocity $v_{I}$ in the laboratory frame, and
an electron have velocity $v_e$ in the rest frame of the ion. Further,
let $\theta$ be the angle between $v_{e}$ and the beam
direction. Hence, $v_{e}^{\parallel}=v_{e}\cos\theta$ is the electron
velocity component parallel to the ion beam. Any electrons that
satisfy $v_I+ v_e^{\parallel}< 0$ will appear in the backward
direction in the laboratory frame. Let the maximum momentum of the
electrons in the ion rest frame be $p_{\rm{max}}$, with
\begin{equation}\label{E:pmax}
p_\mathrm{max} = \sqrt{E_{\rm max}(E_{\rm max} + 2m_e)} \approx \sqrt{2 m_e Q} \,,
\end{equation}
where the approximation holds for small $Q$-values: $E_{\rm max} \approx Q \ll
m_e$. Note that in this regime, ions, as well as electrons,
are non-relativistic and we can use Galilean velocity
addition.\footnote{We have checked explicitly that relativistic
effects are small and can be neglected to good approximation.  Note that
the neutrino is relativistic and its momentum is {\it not} Galilean
invariant.} Let us define an ion momentum $p_I^0$, such that an electron
with momentum $p_\mathrm{max}$ emitted in the backward direction is at
rest:
\begin{equation}\label{E:mom} \frac{p_I^0}{M} =
    \frac{p_\mathrm{max}}{m_e} \quad\text{and}\quad p_I^0 \approx
    \sqrt{\frac{2Q}{m_e}} M \simeq 0.2 \, M \left(\frac{Q}{10\,\rm
    keV}\right)^{1/2} \,.  \end{equation}
From this relation we obtain the boost factor of the ions in the beam.
We have $\gamma = E_I/M \approx \sqrt{1 + 2Q/m_e}$, which is close to
one for the case of interest, when $Q\ll m_e$.  Now one can alter the
ion boost slightly from that value by a small $\Delta p_I = p^0_I - p_I$
such that only electrons close to the endpoint will go in the backward
direction. More precisely, the ion boost will perform a cut such that
only electrons with parallel momentum satisfying
\begin{equation}\label{E:epsilon} p_{\rm{max}}-\epsilon < p_{\parallel}
    < p_{\rm{max}} \end{equation}
will travel in the backward direction in the laboratory frame, where
from Eq.~\ref{E:mom} we get
\begin{equation}\label{E:epsilon2} \epsilon = \frac{m_e}{M} \, \Delta
    p_I \,.  \end{equation}
Hence one selects electrons with momenta close to the endpoint. Note
that fromEq.~\ref{E:epsilon2} the cut in electron momentum $\epsilon$ is
related to the change in the ion momentum by the tiny number $m_e / M$.
Combining this relation with Eq.~\ref{E:pmax} one can translate
$\epsilon$ into a cut on electron energy:

\begin{equation}\label{eq:dp} \Delta E_\beta \approx
    \sqrt{\frac{2Q}{m_e}} \,\epsilon = \frac{\sqrt{2Qm_e}}{M} \, \Delta
    p_I \approx 2Q \frac{\Delta p_I}{p_I^0} \,.  \end{equation}
Therefore, assuming $Q \simeq 5$~keV, sensitivity to
neutrino masses of order $\Delta E_\beta \sim 0.1$~eV requires control of
the ion momentum at the level of $ \delta p/p \sim 10^{-5}$.

\begin{figure}
\begin{center}
\includegraphics[width=0.7\textwidth]{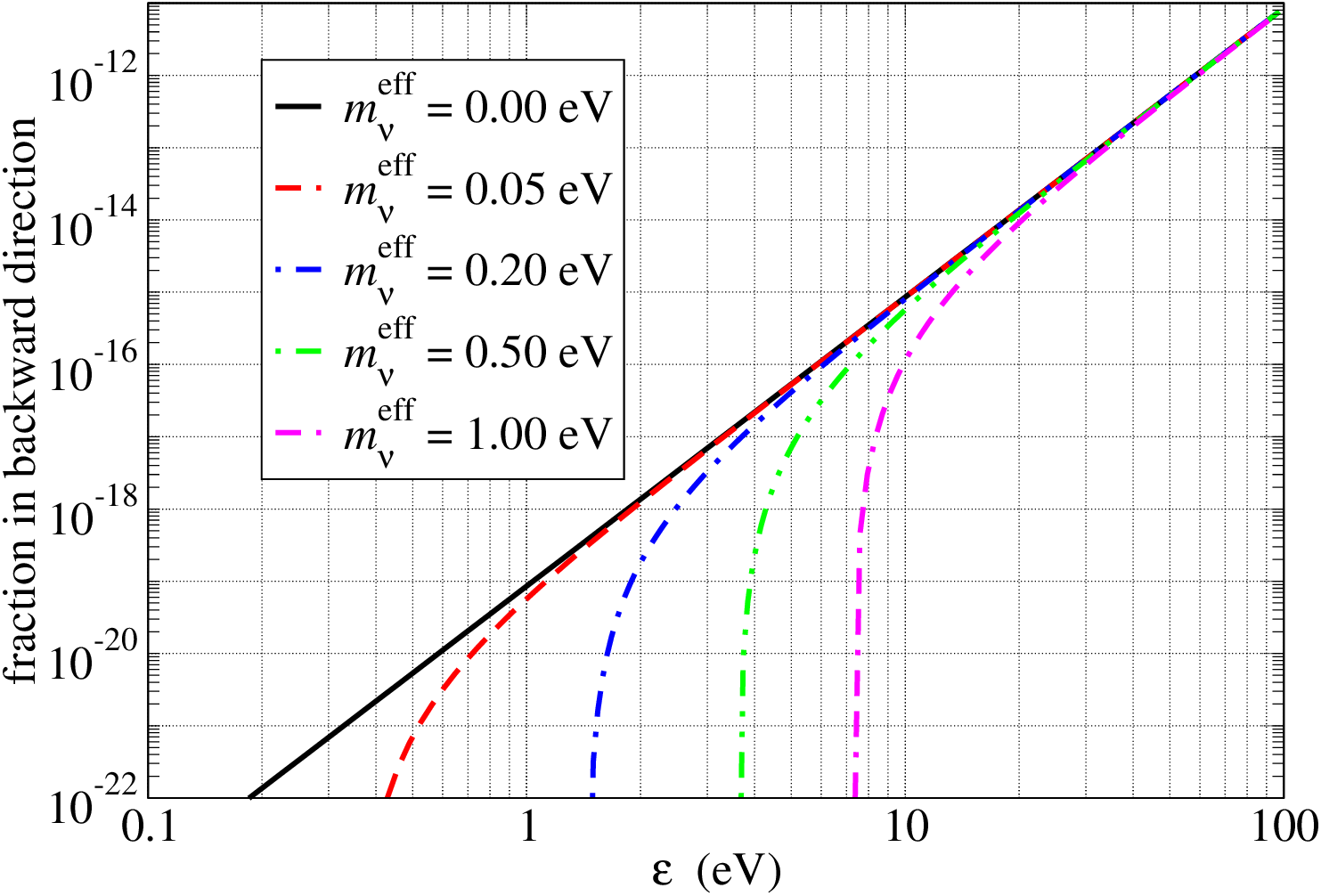}
\end{center}
\mycaption{The fraction of electrons that travel in the backward
direction (Eq.\ref{E:flux}) in the laboratory frame as a function of the
cut on the electron momentum parallel to the beam, $\epsilon$, see
Eq.~\ref{E:epsilon}. A $Q$-value of 5~keV has been considered for a
selection of neutrino masses from zero to $m_{\nu}=1$~eV.
}
\label{Fi:counts}
\end{figure}

The total number of electrons going in the backward direction can be
calculated by transforming to cylindrical coordinates: $p_\perp
dp_\perp dp_\parallel = 2 p (E_\beta + m_e) dE_\beta$ with $p^2 =
p_\perp^2 + p_\parallel^2 = E_\beta(E_\beta + 2m_e)$. Then one finds
\begin{equation}\label{E:flux}
N(\epsilon) = \frac{1}{2}
\int_{p_\mathrm{max} - \epsilon}^{p_\mathrm{max}} dp_\parallel
\int_0^{\sqrt{p_{\rm{max}}^{2}-p_{\parallel}^{2}}} dp_\perp \,
    \frac{d\Gamma}{dE_\beta}\: \frac{p_{\perp}}{p (E_\beta + m_e)}\:.
\end{equation}
In Fig.~\ref{Fi:counts}
we show the number of electrons going backward relative to the total
number of electrons, which is given by 
\begin{equation}
N_\mathrm{tot} = 
2 N(\epsilon = p_\mathrm{max}) =
\int_0^{E_{\rm max}} dE_\beta \, \frac{d\Gamma}{dE_\beta} \,.
\end{equation}
The difference between $m_\nu^\mathrm{eff}=0$ and $m_\nu^\mathrm{eff}
\sim 0.2$~eV is at the level of $10^{-18}$ and we conclude that in
order to reach these sensitivities to the neutrino masses, of order
$\gtrsim 10^{18}$ ion decays are needed. This number, of course,
represents a major challenge for the proposed technique.

Now we assume a magnetic field $B$ parallel to the beam axis. The
electrons will spiral with gyro-radius $p_\perp / qB$ ($q=1$ for
electrons). Hence, electrons with a given orthogonal momentum
component will hit either the front or back wall of the cavern at a
distance from the beam axis ranging from zero to
$R_\mathrm{max}(p_\perp) = 2 p_\perp /B$. The distribution
of events as a function of $R$ encodes some spectral information, the
use of which slightly improves our sensitivity, as described in
Appendix~\ref{S:R}. 

A traditional endpoint search, such as KATRIN uses an electric field
to select electrons close to the endpoint; a cut on the momentum $p >
p_\mathrm{cut}$ such that a shell of the momentum sphere in the ion
rest frame is selected. In our case we cut on the momentum component
parallel to the beam. Hence, we use only a slice of the momentum
sphere, which has the disadvantage that many electrons close to the
endpoint are lost. Roughly, the fraction of electrons in the slice
($p_\mathrm{max} - \epsilon < p_\parallel < p_\mathrm{max}$) and the
shell ($p_\mathrm{cut} = p_\mathrm{max} - \epsilon < p <
p_\mathrm{max}$) is given by $(p_\perp^\mathrm{max} /
p_\mathrm{max})^2 \sim \epsilon / \sqrt{m_e Q} \sim 10^{-4}$, where
$p_\perp^\mathrm{max} \approx \sqrt{2 p_\mathrm{max}\epsilon}$ and the
value $10^{-4}$ is obtained for $\epsilon \sim 5$~eV and $Q\sim
5$~keV. This small factor of useful decays has to be compensated by
the total number of ion decays, which is part of the reason our
proposal requires $> 10^{18}$ decays.


\section{Simulations and sensitivity estimations}
\label{S:analysis}

\subsection{Description of the analysis}
\label{sec:description}

To extract the neutrino mass from an experiment as suggested
here, one would perform a measurement at many values of $\epsilon$ in
order to sample the spectrum shown in Fig.~\ref{Fi:counts}. Since the
count rate increases very fast with $\epsilon$ one would optimize the
measurement time at a given $\epsilon$ such that most time is spent
close to the endpoint at small $\epsilon$, whereas the necessary time
in order to accumulate enough events decreases fast with
increasing $\epsilon$. For each experimental run at fixed $\epsilon$,
one can perform a fit to the distribution of events as a function of
$R$, the distance of the detected event from the beam axis, as
discussed in Appendix~\ref{S:R}. Before commenting on such a
full-fledged analysis, we discuss first a simplified analysis using
only the total number of counts (no radial information) at two values
of $\epsilon$.

At least two data points are needed to simultaneously extract the
$Q$-value of the ion. Typical uncertainties on $Q$ are too large
compared to
the precision required in order to use the $Q$-value as an input for
the endpoint measurement~\cite{Q}. For example, the $Q$-value for
Tritium decay has been determined as $18589.8 \pm 1.2$~eV~\cite{H3Q},
to be compared to the 0.2~eV sensitivity goal of
KATRIN. Future high precision measurements at the MPIK/UW-PTMS
  in Heidelberg~\cite{Blaum} aim at a precision for $Q$ of 30~meV,
  which may be used as cross check for the KATRIN experiment. In
general it is therefore necessary to fit for the $Q$-value, in
addition to the neutrino mass~\cite{Osipowicz:2001sq, Q}.
Even though the neutrino mass and $Q$-value are not
degenerate if full spectral information is available, in an
analysis of the total counts at a single $\epsilon$, the effect of a
non-zero neutrino mass can mimic a slightly larger
$Q$-value. Hence, a measurement at a single value for the ion
boost is not sufficient to fit for the neutrino mass.

In the simulations carried out in this study, we combine two
experimental runs: one with small $\epsilon$ close to the endpoint
and one with large $\epsilon$. At large
$\epsilon$ the total count rate is several orders of magnitude larger
than the change invoked by a non-zero neutrino mass. Hence, a
2-parameter fit for the neutrino mass and $Q$-value at large $\epsilon$
will be largely independent of the neutrino mass, i.e.\ one
effectively makes a measurement of the $Q$-value. For small
$\epsilon$, however, the total count rate becomes comparable to the
reduction for non-zero neutrino mass.  There is a strong neutrino mass
dependence in this case which, when combined with the $Q$-value
measurement from the run with large $\epsilon$, constrains the
neutrino mass.

Quantitatively, in our numerical analyses we calculate the following $\chi^{2}$
\begin{equation}\label{E:chi_min}
\chi^{2}=\mineta\left[\chi^{2}_{\epsilon_1}+\chi^{2}_{\epsilon_2} 
    + \left(\frac{\eta}{\sigma_\mathrm{norm}}\right)^{2}\right].
\end{equation}
Our $\chi_{\epsilon_i}^{2}$ definition is based on Poisson statistics:
\begin{equation}\label{E:chi}
\chi^{2}_{\epsilon_i}=2\;\left[N_{\epsilon_i}-D_{\epsilon_i} 
      +D_{\epsilon_i}\;
     \ln\left(\frac{D_{\epsilon_i}}{N_{\epsilon_i}}\right)\right]
\end{equation}
where 
\begin{equation}
    N_{\epsilon_i}=(1+\eta)\, T_{\epsilon_i}(m^{\rm eff}_\nu, \Delta Q) 
\,,\qquad
D_{\epsilon_i} =
T_{\epsilon_i}(m_\nu^\mathrm{true}, \Delta Q = 0) \,.
\end{equation}
$T_{\epsilon_i}(m^{\rm eff}_\nu, \Delta Q)$ is the predicted number of
events for displacement $\epsilon_i$ for a certain neutrino mass
hypothesis $m^{\rm eff}_\nu$ and a shift in the $Q$-value of $\Delta Q$.  The
pull parameter $\eta$ takes into account a systematic uncertainty on
the overall normalization of the number of events,
$\sigma_\mathrm{norm}$, which we assume to be correlated between the
two runs. $D_{\epsilon_i}$ is the simulated data for displacement
$\epsilon_i$ at an assumed ``true value'' for $m^{\rm eff}_\nu$ and at the
``true'' $Q$-value. The final
$\chi^{2}$ is found by adding the penalty term $(\eta/\sigma_{\rm
norm})^2$ and then minimizing with
respect to the pull. Owing to the large uncertainties on the
$Q$-value, we always treat $\Delta Q$ as a free parameter.

\begin{figure}
\begin{center}
\includegraphics[width=0.95\textwidth]{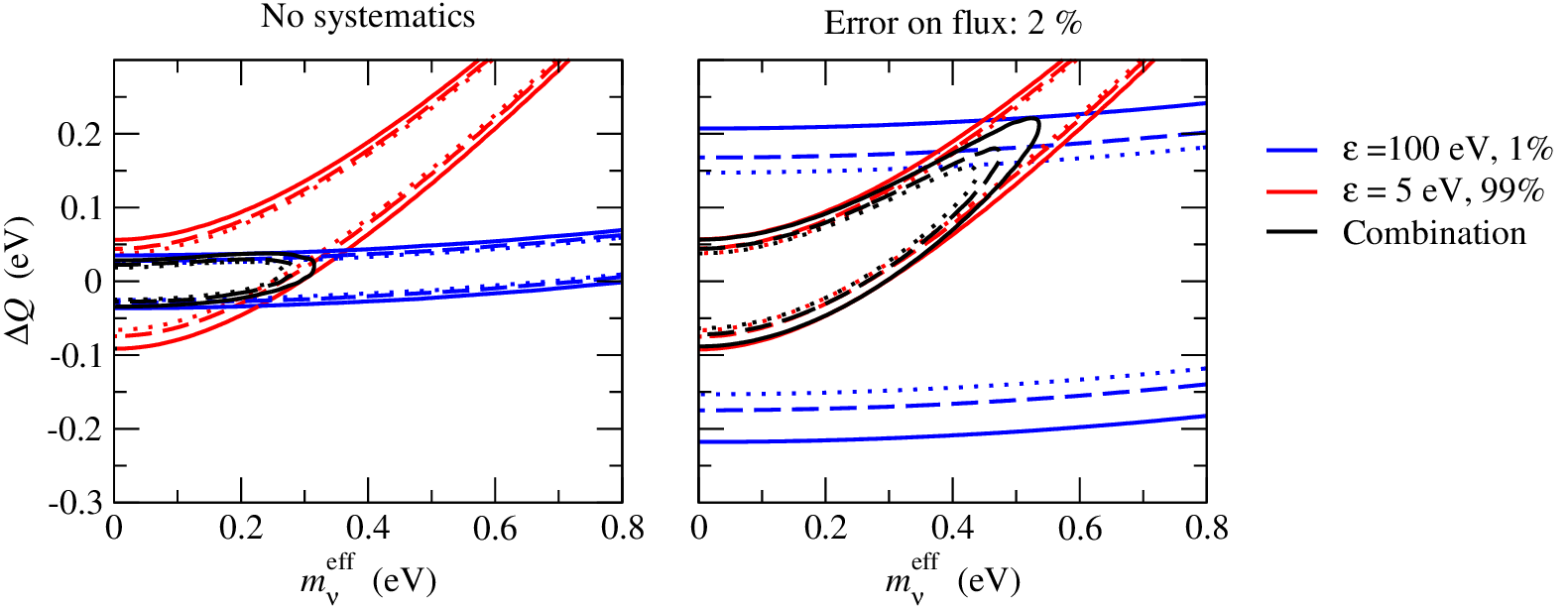}
\end{center}
\mycaption{We show the 90\%, 95\% and 99\% confidence level contours for
  two degrees of freedom. We assume a $Q$-value of 5~keV, $10^{18}$
  useful ion decays, and a ``true'' neutrino mass of
  $m_{\nu}=0.1$~eV. Results are shown separately for $\epsilon = 5$~eV
  and $\epsilon = 100$~eV, and their combination. In the left panel no
  systematical errors are assumed, whereas in the right panel, we
  introduce a 2\% error on the flux normalization. No backgrounds have
  been included in these analyses.
}
\label{Fi:2parfits} 
\end{figure}

In Fig.~\ref{Fi:2parfits} we demonstrate this approach for the case
$Q=5$~keV and $m_{\nu}^\mathrm{true}=0.1$~eV, both for no systematics
and a 2\% error on the normalization of the flux. Runs with $\epsilon
=5$~eV and 100~eV have been taken with measurement times in the ratio
99:1 such that the total number of useful ion decays is $10^{18}$. It
is evident from Fig.~\ref{Fi:2parfits} that for large $\epsilon$, the
fit only has a very slight dependence on the neutrino mass and
provides a measurement of $Q$, where the accuracy on the $Q$-value is
sensitive to uncertainty on the normalization of the number of useful
decays. For small $\epsilon$, on the other hand, we see a strong
dependence on the neutrino mass which shows little variation with the
normalization of the flux. Obviously, a low $\epsilon$ measurement
alone cannot constrain $m_\nu^\mathrm{eff}$ because of the correlation
with $\Delta Q$. However, the combination of small and large
$\epsilon$ runs provides sensitivity to the neutrino mass. The effect
of the systematics on the neutrino mass sensitivity is felt through
the uncertainty on the $Q$-value.

We have checked that the two-$\epsilon$ run analysis provides already
a good sensitivity. Certainly there is room for improvement by
exploring many runs at different $\epsilon$ values with optimized measurement
times at each position. While a detailed optimization along these
lines is beyond the scope of the present work, preliminary studies
indicate that the sensitivity may be improved by a factor of order
10\%, similar to exploring the radial information as described
Appendix~\ref{S:R}.

Fig.~\ref{Fi:2parfits} indicates that the proposed measurement may
provide a determination of the $Q$-value at the sub-eV level. Let us
stress that this would rely on a perfect knowledge of the momentum of
the ion in the beam $p_I$. It turns out that $\Delta Q$ is fully
correlated with $p_I$ and what actually is measured is the sum of the
two. Therefore, if we are interested mainly in a measurement of the
neutrino mass, without the ambition to determine also $Q$, we do not
need to know $p_I$ with very good accuracy, since we simply fit for
the sum $p_I+Q$. However, it is very important that the {\it momentum
  spread} $\delta p_I$ of the beam is small enough, such that the
spectral distortion introduced by $m_\nu^\mathrm{eff}$ is not washed
out. We are going to quantify this requirement below.

\subsection{Requirements to reach sub-0.2 eV sensitivity on the neutrino mass}
\label{sec:requirements}

\begin{figure}
\begin{center}
\includegraphics[width=0.7\textwidth]{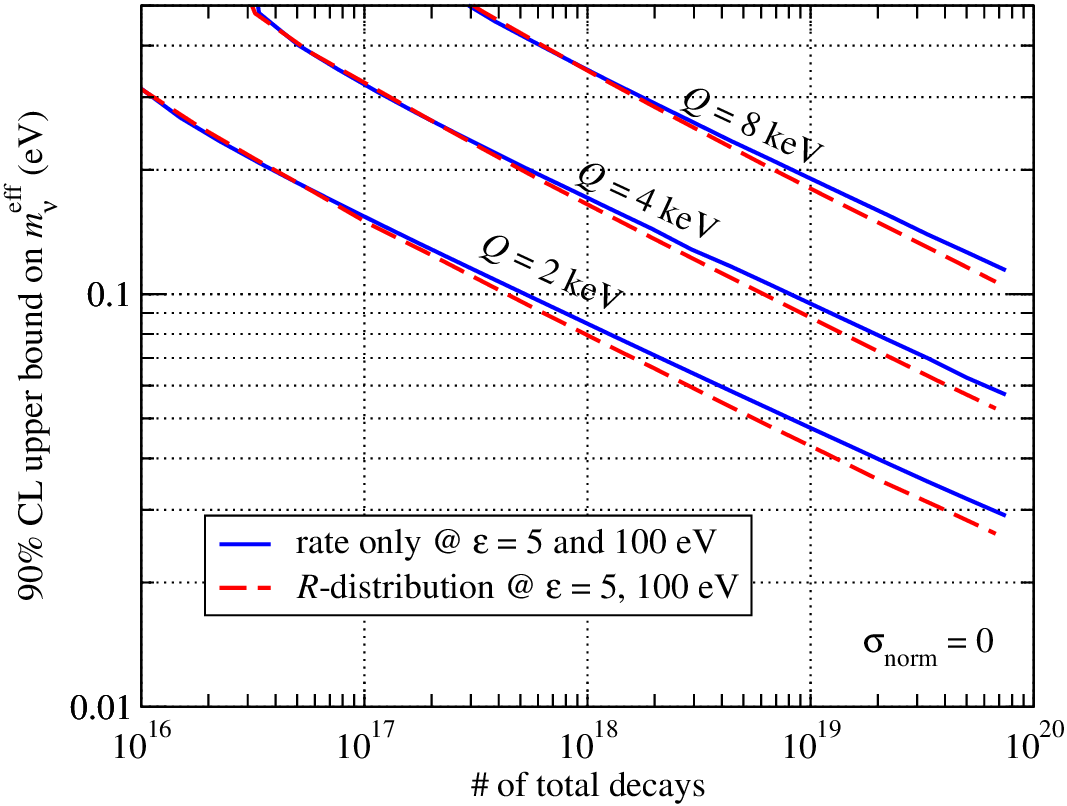}
\end{center}
\mycaption{Sensitivity to the neutrino mass at 90\% confidence level
  as a function of useful decays for $Q=2,4,8$~keV. We show
  the total rate analyses with two ion boosts corresponding $\epsilon
  = 5$ and 100~eV (solid) and an analysis with using also the radial
  distribution with 20 bins of equal width in $R$
  (dashed).}\label{Fi:sens_events}
\end{figure}

In Fig.~\ref{Fi:sens_events}, we present the upper bound on $m^{\rm
  eff}_\nu$ at 90\% confidence level, which can be obtained if the
true value is $m^{\rm true}_\nu = 0$, as a function of useful decays
for $Q$-values of $2$, $4$ and $8$~keV. Runs of $\epsilon = 5$~eV and
$\epsilon = 100$~eV are considered in the ratio 99:1 such that the
total useful number of decays sum up to the value shown on the
horizontal axis. No backgrounds or systematics have been included, and
we neglect the momentum spread of the ions. It is seen from
Fig.~\ref{Fi:sens_events} that the approach adopted in this paper
places very strong requirements on $Q$ and the luminosity. To match
the expected sensitivity of the KATRIN and MARE experiments of
$m_{\nu}^{\rm eff}<0.2$ eV, one requires $4\cdot 10^{16}$, $5\cdot
10^{17}$, $8\cdot 10^{18}$ decays for $Q$-values of 2, 4, 8~keV,
respectively. A factor 5 improvement down to $m^{\rm
  eff}_\nu<0.04$~eV, which will separate the normal and inverted mass
hierarchy regions in Fig.~\ref{Fi:effmass}, requires in excess of
$10^{19}$ counts across the run of the experiment.

We also show in Fig.~\ref{Fi:sens_events} an analysis that takes into
account the radial
distribution of the backward moving electrons, as described in
Appendix~\ref{S:R}. 20 bins of equal radial width
were used with the same run parameters as the total rates
analysis. There is only minor improvement when radial information is
included in the very high luminosity region, because this approach
does not reconstruct the electron energy spectrum itself, as electrons
with transverse momentum $p_{\perp}$ can strike the detector at any
radius less than twice their gyro-radius $r_{g}=2p_{\perp}/B$. The
spectral information is thus significantly smeared out and cannot be
reconstructed accurately. Although neither analysis has been
rigorously optimized, one does not expect order of magnitude increases
in sensitivity. Therefore, in the following, we only consider the
two-$\epsilon$ run analysis described in section~\ref{sec:description}.

\begin{figure}
\begin{center}
\includegraphics[width=0.9\textwidth]{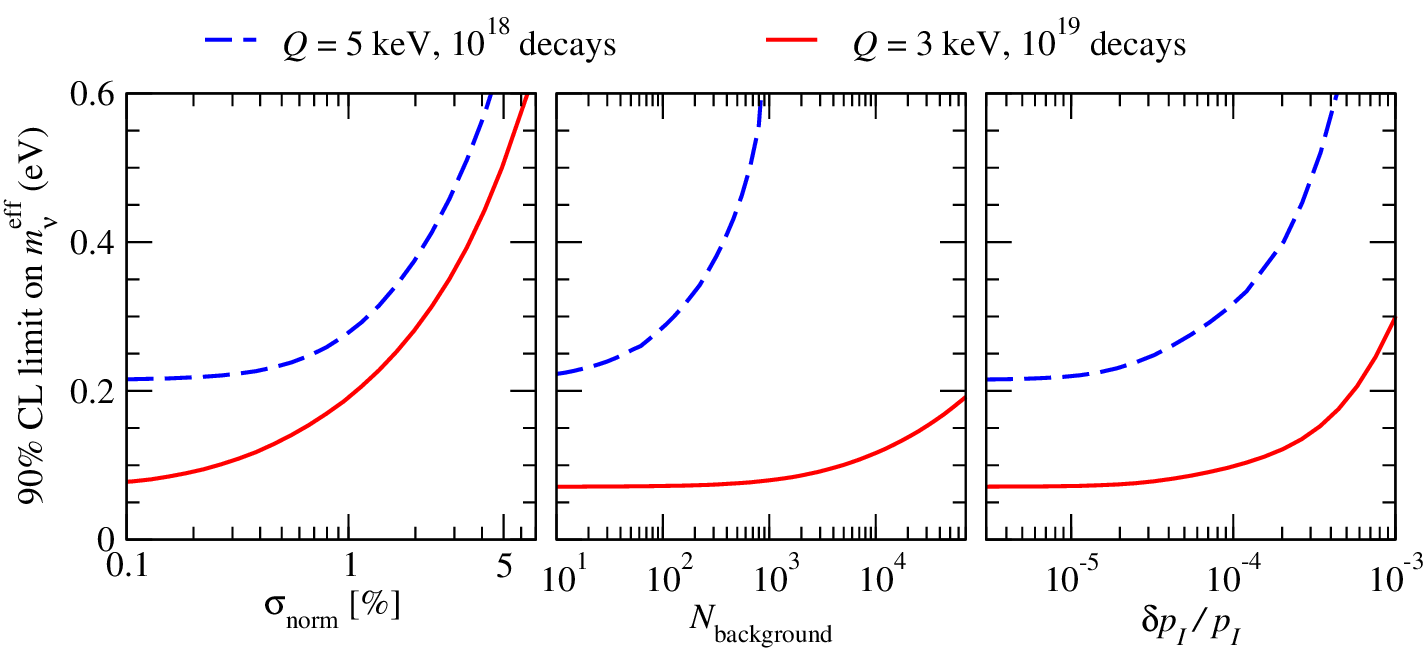}
\end{center}
\mycaption{Sensitivity to $m^{\rm eff}_\nu$ at 90\% confidence level for
  $Q=5$~keV and $10^{18}$ useful decays (dashed/blue) and $Q=3$~keV
  and $10^{19}$ useful decays (solid/red). In the left panel, the
  effect of the normalization uncertainty on the flux is considered;
  background levels are varied in the center panel; whilst the effect
  of the momentum spread of the initial ion is taken into account in
  the right panel.}
\label{Fi:systematics}
\end{figure}

In Fig.~\ref{Fi:systematics} we examine the behavior of
neutrino mass sensitivity with systematics, backgrounds and the ion
beam momentum spread. We consider the cases $Q=3$~keV with $10^{19}$
useful decays (``high sensitivity'') and $Q=5$~keV with $10^{18}$
useful decays (``low sensitivity''), and, as before, in both cases we
run at $\epsilon = 5$ and 100~eV with a ratio of 99:1. The ``low'' and
``high'' sensitivity configurations provide nominal sensitivities of
0.21 and 0.071~eV, respectively. Now we include each one of the three
above mentioned effects separately, in order to investigate at which
level the sensitivity starts to deviate from these idealized numbers.

From the left panel of Fig.~\ref{Fi:systematics}, it is seen that the
sensitivity starts to deteriorate rapidly for normalization
uncertainties greater than 1\%. This can be understood in conjunction
with Fig.~\ref{Fi:2parfits} where it was shown how the error on the
normalization on the number of useful decays affects the overall
sensitivity. The systematics severely limits the ability of the
$\epsilon =100$~eV run to constrain the $Q$-value which in turn
worsens the sensitivity to the neutrino mass. This is true for both
cases considered, but especially the high sensitivity setup which
becomes independent of $\sigma_\mathrm{norm}$ only below
0.1\%. 
Naively such an accuracy seems not very difficult to
achieve. Apart from traditional beam flux determinations, 
one could measure the electrons emitted in the forward
direction by installing an electron detector at the
front wall of the chamber. The $\mathcal{O}(10^{18})$ electrons may provide a
means to determine the total number of decays with negligible
statistical error.

In the center panel of Fig.~\ref{Fi:systematics}, the effect of
including backgrounds is shown. We assume that the magnetic field is
adjusted such that the total area illuminated
by the electrons on the back wall is the same for both the small and
large $\epsilon$. A constant background rate is taken. 
The number of background events shown on the horizontal
axis is the combined background from both runs. Here, the behavior
is as one would expect; the low sensitivity setup suffers badly with
loss in sensitivity even for background rates less than 100 across the
entirety of the experiment. The high sensitivity setup performs much
better; it is able to tolerate up to 1000 background events before the
sensitivity starts to diminish. 

In the proposed experiment there are several potentially dangerous
sources of backgrounds. Even a tiny fraction of the $10^{18}$ to
$10^{19}$ electrons supposed to go in the forward direction, which
could be deflected by some effect to the back wall, would constitute a
serious background for the measurement. Note that the forward-going
decay electrons are high energy, and can eject electrons from the
forward wall. These could then spiral backwards and hit the rear
wall. This could be mitigated somewhat by removing the magnetic field
in the forward direction, and using timing information of the beam
bunch relative to the electron detection. We also ignore the electric
and magnetic field generated by the beam, which will affect the
trajectory of the ejected electrons.  Beam-related backgrounds can be
measured by adjusting the ion's momentum so that $\epsilon < 0$ and no
electrons should have a backward trajectory, while the beam-unrelated
background can be determined from beam-off periods.

In the right panel of Fig.~\ref{Fi:systematics}, we present the
sensitivity as a function of the spread of the ion momentum in the
beam direction. We assume a Gaussian momentum distribution in the beam
direction with width $\delta p_I$. From Eq.~\ref{eq:dp}, this
translates into a spread on $\epsilon$ of
\begin{equation}
\delta\epsilon = \sqrt{2Qm_e} \, \frac{\delta p_I}{p_I} \,.
\end{equation}
In order to include this uncertainty in the analysis we fold the
number of events $N(\epsilon)$, given in Eq.~\ref{E:flux}, with a
Gaussian with width $\delta\epsilon$. This procedure takes into
account the momentum spread parallel to the beam, whereas the momentum
uncertainty perpendicular to the beam is not included.  This will be
important in reality, especially in a setup that uses the radial
distribution of electrons.

In the right panel of Fig.~\ref{Fi:systematics} one can see that for
the low sensitivity setup, the sensitivity to the neutrino mass starts
to abate at $\delta p_{I}/p_{I} \sim 2\cdot 10^{-5}$. This is
consistent with the relation between the uncertainty on the electron energy and
the ion momentum spread: 
\begin{equation}\label{E:uncert}
\delta E_\beta = 2E_\beta\frac{\delta p_{I}}{p_{I}} + \mathcal{O}\left[\left(
    \frac{E_\beta}{m_{e}}\right)^{2}\right] \,.
\end{equation}
For $\delta E_\beta = 0.22$~eV and $Q=5$~keV, one requires an
uncertainty on the momentum to be $\delta p_{I}/p_{I} < 2.2\cdot
10^{-5}$ for $E_\beta \approx Q$.
The same estimate for the high sensitivity setup predicts $\delta
E_\beta \sim 0.075$~eV for $\delta p_{I}/p_{I} \sim 1.2\cdot 10^{-5}$.
However, from Fig.~\ref{Fi:systematics} we observe that the
sensitivity starts to deteriorate only at $\delta p_{I}/p_{I} \sim
1.0\cdot 10^{-4}$ where, according to Eq.~\ref{E:uncert}, $\delta E_\beta
\sim 0.6$~eV. This behaviour indicates that the higher count rates can
compensate for less precision on the ion beam momentum. 
In summary, to reach interesting sensitivities to the neutrino mass,
momentum spreads in the range $10^{-5}$ to $10^{-4}$ are required.


\section{Challenges and technology requirements}
\label{S:technology}

\subsection{Total number of decays}

Our analysis indicates that we would require $10^{18} - 10^{20}$
useful decays to reach sub-KATRIN sensitivities.  Such production is
in line with projected EURISOL intensities~\cite{eurisolintens,
 Volpe:2003fi}. However, to store such a large number of ions is a
major challenge, especially as they will be long lived and possibly
have high charge.  Space charge effects, the defocussing of the beam
due to beam-particle interactions, will be significant at
non-relativistic velocities.

\subsection{Momentum spread requirements and coulomb crystals in storage rings}

The mass measurements discussed in this paper can only be realized if
the energy spread of the ions in the ring is very small. For an
electron with kinetic energy $E_\beta$, an uncertainty on the ion
momentum $\delta p_{I}/p_{I}$ translates to an uncertainty on the
electron energy via Eq.~\ref{E:uncert}.  For example, an endpoint
electron from an ion with $Q=5$~keV, an uncertainty on its kinetic
energy of 0.1~eV requires $\delta p_{I}/p_{I} = 1\cdot 10^{-5}$. The
numerical results presented in section~\ref{sec:requirements} indicate
that in order to reach interesting sensitivities to the neutrino mass,
ion momentum spreads in the range $10^{-5}$ to $10^{-4}$ are required.
Values in this range may be achieved with classical beam cooling for
low energy ion beams with electron cooling or laser cooling.
Therefore, if ions with very low $Q$-values ($\mathcal{O}(1 \, {\rm
  keV})$) are available, classical cooling is sufficient.
Availability of such low $Q$-values in nature is minimal, however,
those that do exist are not practical. For example, $^{187}$Re has
$Q=2.6$~keV, but it also has a half-life of $4.5\cdot 10^{10}$ years.
New coolings techniques will be required for the likely much
larger $Q$-values. 

For cooled low intensity ion beams of $5000-10 000$ ions, a distinct
transition to much lower momentum spread has been observed with
increasing electron cooling current at NAP-M in 
Novosibirsk~\cite{par84}, ESR at GSI~\cite{ste96} and CRYRING at 
MSL~\cite{dan02}.
This has been interpreted as an ordering of the ions in a regime where
the energy spread of the beam is too small to permit any individual
ion to overtake or drop behind its neighbors. Effectively, the ions
will create a one dimensional string of ions with a minimum distance
separating individual ions. This distance is defined classically by
setting kinetic and potential energy equal, giving
$d_{min}=(Zq)^2/4\pi\epsilon_0kT$ where $kT$ is the ion temperature
\cite{dan02}. The momentum spread is measured to be smaller than
$10^{-6}$ with the upper limit given by the known ripple of the power
supplies controlling the electromagnetic fields which confine the
beam.  The ultimate momentum spread is set by the cooling forces
acting on the beam. First studies have shown that bunching of ordered
beam should be possible enabling at least moderate acceleration
\cite{dan03}. To create much higher energy ordered beams it is
necessary to develop cooling schemes for highly relativistic beams as
the ordered beam bunch is unlikely to survive transition. Development
of electron cooling schemes for ultra relativistic ion beams are
underway at BNL~\cite{bnl-HE-cooling}, for example. The new HE-storage
rings at the FAIR facility in Germany will in time enable
experiments with laser cooling of highly relativistic 
beams~\cite{fair-laser-cooling}.

For the low energy regime, experimental work is in progress in Japan at
the S-LSR ring in Kyoto \cite{nod06}. They have demonstrated one
dimensional ordering of protons \cite{shi07} and are working on special
tapered cooling schemes in a dispersion free storage ring for
experiments aiming to achieve three dimensional ordering of highly
charged ions. 

Potential new applications such as the one proposed in this paper could
motivate R\&D on ordered beams. Such efforts could eventually yield 
higher intensity and/or higher energy ordered beams.  Studies are needed
on increasing the number of ions, and ensuring the desired properties of
the beam at high densities.  All experiments on crystallized beams
carried out thus far \cite{schramm04} have used less than 10000 ions.

\subsection{Comments on $Q$-values}

In this paper, we have established the necessity of $> 10^{18}$ decays
and $Q$-values $\sim 1-10$~keV if accelerated ions are to be used to
measure the neutrino mass at sub-KATRIN levels. This measurement
therefore needs a low $Q$-value ion with a short half-life.
Rhenium-187 possesses the smallest $Q$-value known: 2.6 KeV; however,
the $5/2^{+}\rightarrow 1/2^{-}$ transition has a half-life of
$4.5\cdot 10^{10}$~y. For the set of beta emitting nuclei with
half-lives $< 10$~y, Ruthenium-106 and Radium-228 have the smallest
$Q$-values: 39.4~keV and 45.9~keV, respectively. The total number of
decays will need to be increased by 3-4 orders of magnitude to match
the sensitivities discussed in Sec.~\ref{sec:requirements} if these
ions were to be used.

The most promising ion is therefore Tritium with $Q=18.6$~keV, however
electron cooling is necessary rather than laser cooling because it has
no hyperfine structure when ionized. Tritium has the added advantage
that the number of ions that can be obtained is orders of magnitude
larger than those discussed in this paper, as it is naturally occurring
and does not need to be produced in an accelerator.


\section{Summary and Conclusions}
\label{S:summary}

We have discussed the possibility to use radioactive ion beams to study
the kinematic effects of a non-zero neutrino mass close to the
endpoint of the beta decay spectrum. The underlying idea is to adjust the
boost factor of the ions in such a way that only electrons close to
the endpoint will have backward trajectories in the laboratory frame. 
We have discussed some
requirements of the proposed setup in order to exceed
the sensitivity of 0.2~eV of the latest generation of Tritium and
Rhenium decay experiments. A crucial question is whether it will be
possible to accelerate enough ions within reasonable time such that of
order $10^{18} - 10^{20}$ decays can be observed. This issue is also related
to the identification of a suitable ion with a low enough
$Q$-value (in order to maximize the effect of the neutrino mass), and a
small enough half life (in order to have a high enough decay rate).

As an example, for a very low $Q$-value of 2~keV, one needs $4\cdot
10^{16}$ decays to obtain the KATRIN sensitivity of 0.2~eV, while
$10^{19}$ decays will allow for a $m^{\rm eff}_\nu$ measurement at 0.04~eV. On
the other hand, if no suitable ion with such a low $Q$-value can be
identified the requirements on the total number of decays increases
drastically: for $Q = 4 \, (8)$~keV the 0.2~eV sensitivity is reached
for $5\cdot 10^{17}$ ($8\cdot 10^{18}$) decays, respectively.  The
sensitivity goal of $m^{\rm eff}_\nu<0.04$~eV, which will separate the
normal and inverted neutrino mass hierarchy regions, requires in
excess of $10^{19}$ counts across the run of the experiment together
with a $Q$-value of 2~keV.

Our proposal relies on a very small momentum spread of the ions, at
the level of $\delta p_I/p_I \lesssim 10^{-5}$.  Momentum spreads in
the range $10^{-4}$ to $10^{-5}$ may be reached by classical beam
cooling techniques such as electron or laser cooling. For even smaller
momentum spreads one may explore the use of ``crystallized
beams''~\cite{crystallized, schramm04}. This is a phenomenon which has
been observed for cooled low intensity ion beams, resulting in a
transition to an ordered state of the beam with momentum spreads of
less than $10^{-6}$. It has yet to be proven whether this technique
allows for the extremely high beam intensities required for the
neutrino mass measurement proposed here.
In addition one needs to separate forward and backward moving
electrons with an precision of order $10^{-16}$. In general forward
moving electrons will create backward moving electrons by collisions
with the walls of the cavern. A suitable experimental configuration to
overcome this challenge must be identified.

Given the utmost importance of establishing the absolute value of the
neutrino mass in a model independent way, we feel that all possible
directions have to be explored. We hope that our work will stimulate
more intense investigations on the possibility to use ion beams to
pursue this fundamental issue in neutrino physics.


\acknowledgments

The authors give thanks for the many useful comments and discussions had
with colleagues throughout this work: Jose Bernabeu, Klaus Blaum, Michel
Doser, Adrian Fabich, Yuri Litvinov, Mauro Mezzetto, Yu Novikov,
Silvia Pascoli, Karsten Riisager and Paul Soler.
CO acknowledges the support of a STFC studentship and CARE, under
contract number RII3-CT-2003-0506395, for overseas fieldwork
support. CO would also like to thank the ISOLDE group and beta beam
group at CERN for their hospitality on a number of occasions.  TS
acknowledges support by the Transregio Sonderforschungsbereich TR27
``Neutrinos and Beyond'' der Deutschen Forschungsgemeinschaft. ML, CO
and TS acknowledge the financial support of the European Community
under the European Commission Framework Programme 7 Design Study:
EUROnu, Project Number 212372. The EC is not liable for any use that
may be made of the information contained herein.

\appendix

\section{Radial distribution}
\label{S:R}

In this appendix we comment on the radial distribution of the electron
events on the back wall of the chamber. One could use a silicon pixel
detector which gives position information.  One can bin the events as
a function of the distance from the beam and perform a fit to the $R$
distribution for each experimental run at given momentum cut
$\epsilon$.

\begin{figure}
\begin{center}
\includegraphics[width=0.3\textwidth]{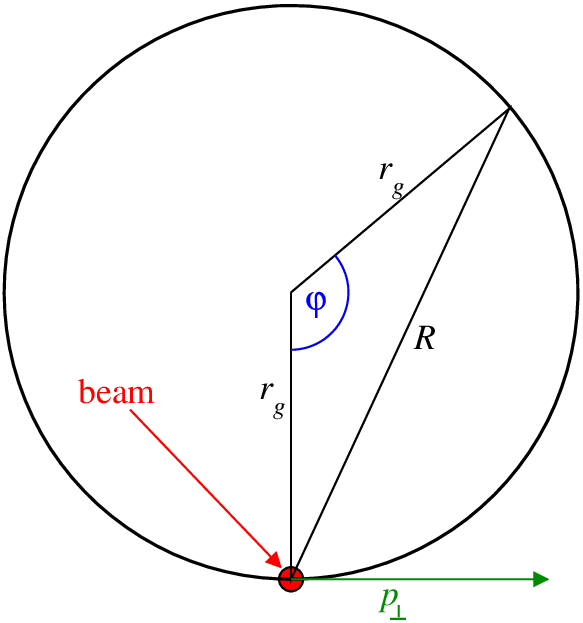}
\end{center}
\mycaption{Trajectory of an electron in the plane perpendicular to the
  beam.}
\label{Fi:R}
\end{figure}

Consider an electron with charge $q = 1$ and momentum $p_{\perp}$
perpendicular to a magnetic field $B$, which is parallel to the beam
axis. The trajectory of the particle will be a helix with the maximum
distance from the beam axis given by twice the gyro-radius
\begin{equation}
R_{\rm{max}}=2r_{g}=\frac{2p_{\perp}}{qB} \,.
\end{equation}
The trajectory of the electron in the plane orthogonal to the beam is
shown in Fig.~\ref{Fi:R}, where $R$ is the distance from the
beam.  The electrons will hit the detector with a flat
distribution in the angle $\varphi$.  We obtain for the differential count
rate (compare Eq.~\ref{E:flux}):
\begin{equation}
dN = \frac{1}{2} \,
    \frac{d\Gamma}{dE_\beta}\, \frac{p_{\perp}}{p (E_\beta + m_e)}\,
    dp_\parallel dp_\perp \, \frac{d\varphi}{2\pi} \,.
\end{equation}
From Fig.~\ref{Fi:R} one finds
\begin{equation}
R = 2 r_g \sin\frac{\varphi}{2} \quad\text{and}\quad
d\varphi = dR \, \frac{B}
{p_\perp \sqrt{1 - \left(\frac{R B}{2p_\perp}\right)^2}} \,.
\end{equation}
Hence, we obtain
\begin{equation}
dN = \frac{B}{4\pi} \,
    \frac{d\Gamma}{dE_\beta}\, \frac{1}{p (E_\beta + m_e)}\,
    \frac{1}{\sqrt{1 - \left(\frac{R B}{2p_\perp}\right)^2}} \,
    dp_\parallel dp_\perp \, dR \,.
\end{equation}
Integration of the momentum for a given $\epsilon$ yields the $R$
distribution of the events:
\begin{equation}\label{E:dNdR}
\frac{dN}{dR} = \frac{B}{4\pi} \,
    \int_{p_\perp^\mathrm{min}(R)}^{p_\perp^\mathrm{max}} 
    dp_\perp \,
    \frac{1}{\sqrt{1 - \left(\frac{R B}{2p_\perp}\right)^2}} 
    \int_{p_\mathrm{max} - \epsilon}^{p_\parallel^\mathrm{max}(p_\perp)}
    dp_\parallel \,
    \frac{d\Gamma}{dE_\beta}\, \frac{1}{p (E_\beta + m_e)}\, ,
\end{equation}
where $p_\perp^\mathrm{max} = \sqrt{2 p_\mathrm{max} \epsilon -
  \epsilon^2}$ is the maximum perpendicular momentum for given
$\epsilon$, $p_\perp^\mathrm{min}(R) = BR / 2$ is the minimum
perpendicular momentum required to hit at a distance $R$ from the
beam, and $p_\parallel^\mathrm{max}(p_\perp) = \sqrt{p_\mathrm{max}^2
  - p_\perp^2}$. Note that integrating Eq.~\ref{E:dNdR} over $R$
from zero to $R_\mathrm{max}$ returns Eq.~\ref{E:flux}.

\begin{figure}
\begin{center}
\includegraphics[width=0.5\textwidth]{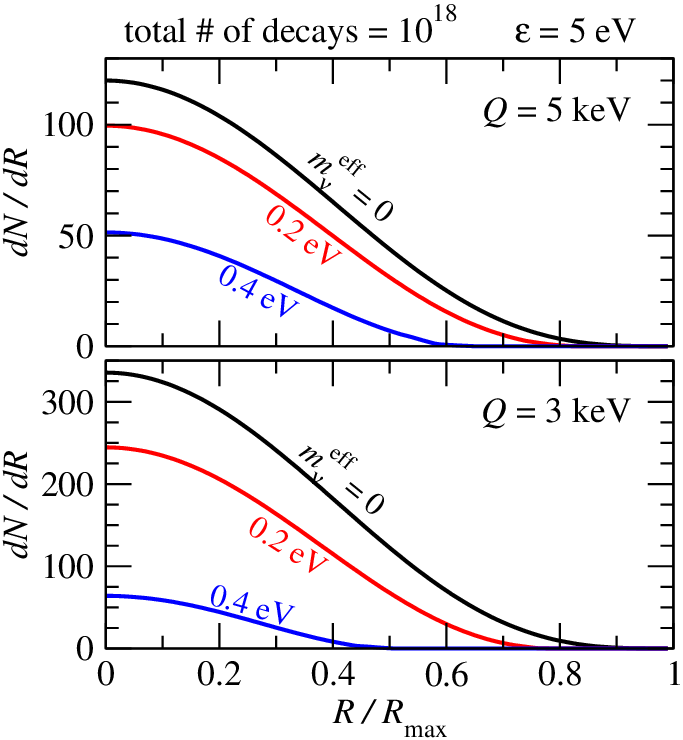}
\end{center}
\mycaption{Number of events as a function of the distance $R$ from the
  beam axis for $10^{18}$ decays at an ion boost corresponding to
  $\epsilon = 5$~eV. The distribution is shown for $Q = 3$ and 5~keV
  and for $m_\nu^\mathrm{eff} = 0, 0.2, 0.4$~eV.}
\label{Fi:dN_dr}
\end{figure}

In Fig.~\ref{Fi:dN_dr} we show some examples for the $R$
distribution. Unfortunately the shape of this distribution carries
only limited information on the neutrino mass. The reason is that
there is a significant averaging of the endpoint
region. Note that the electrons with $p_\perp^\mathrm{max}$, which
have the maximal momentum $p_\mathrm{max}$, will hit the wall at all
radii from zero to $R_\mathrm{max}$. Hence, for small $R$ electrons
with rather broad range of momenta contribute. Because of this the
$R$ distribution provides only limited additional information, beyond
just the decrease of event number with increasing $m_\nu^\mathrm{eff}$. This
explains the only modest improvement of the sensitivity seen in
Fig.~\ref{Fi:sens_events}.


\end{document}